\documentclass{article}

\usepackage{amsmath,amssymb}
\newcommand{\ket}[1]{| #1 \rangle}
\newcommand{\bra}[1]{\langle #1 |}
\newcommand\ad{{\bf \hat{a}_{1}^{\dag}}}
\newcommand\aponto{{\bf \hat{a}_{1}\cdot}}

\newcommand\pad{{\bf \cdot\hat{a}_{1}^{\dag}}}
\newcommand\bd{{\bf \hat{a}_{2}^{\dag}}}
\newcommand\bp{{\bf \hat{a}_{2}\cdot}}
\newcommand\bdp{{\bf \hat{a}_{2}^{\dag}\cdot}}

\newcommand\pbd{{\bf \cdot\hat{a}_{2}^{\dag}}}
\newcommand\cmat{{ m}_{1}{(t)}}
\newcommand\ccmat{{ m}^{*}_{1}{(t)}}
\newcommand\cmbt{{ m}_{2}{(t)}}
\newcommand\ccmbt{{ m}^{*}_{2}{(t)}}

\newcommand\cjat{{ j}_{1}{(t)}}
\newcommand\cjbt{{ j}_{2}{(t)}}
\newcommand\czt{{ z}{{(t)}}}
\newcommand\cczt{{ z}^{*}{{(t)}}}

\newcommand\cqt{{ q}{{(t)}}}
\newcommand\ccqt{{ q}^{*}{{(t)}}}

\newcommand\pra{\textit{Phys. Rev. A}}
\newcommand\prb{\textit{Phys. Rev. B}}
\newcommand\prl{\textit{Phys. Rev. Lett.}}

\begin{document}

\title{Decoherence Sturdy
Memories in the Presence of Quantum Dissipation
}

\author{ K.~M. Fonseca Romero$^1$, S.~G. Mokarzel$^{2,3}$, 
and M.~C. Nemes$^{2,4}$\\ 
$^1$ Departamento de F\'\i sica, Universidad Nacional, Bogot\'a,
Colombia\\
$^2$Departamento de F\'{\i}sica--Matem\'atica, Instituto de F\'{\i}sica, Universidade de S\~ao Paulo,\\
 CP 66318, 05315-970 S\~ao Paulo, S.P., Brazil \\ 
$^3$ Departamento de F\'{\i}sica, Pontificia Universidade Cat\'olica
de S\~ao Paulo,\\ R. Marqu\^es de Paranagu\'a, 111, S\~ao Paulo, S.~P., Brazil\\
$^4$ Departamento de
F\'{\i}sica, ICEX, Universidade Federal de Minas Gerais, \\Belo
Horizonte, M.G., Brazil \\}
\maketitle

\begin{abstract}

We consider N identical oscillators coupled to
a single environment and show that the conditions for the existence of
decoherence free subspaces are degeneracy of the oscillator
frequencies and separability of the coupling with the environment. A
formal exact equation for the evolution in the case of two oscillators
is found and the decoherence free subspace is explicitly determined. A
full analytical solution for any initial condition and general
parameters (frequencies and dissipation constants) is given in the markovian
approximation and zero temperature. We find that slight relaxation of
degeneracy and separability conditions leads to the appearence of two
components in the dynamical evolution with very different decoherence
times. The ratio between the characteristic time of the weak and
strong decoherent components is given by $\tau_{WD}/\tau_{SD}\approx k/\delta k$,
where $\delta k$ is a measure of the nonseparability of the coupling to the
environment and $k$ is the mean decay constant of the oscillators.

\end{abstract}




\section{Introduction}

The very same  mechanism responsible for the potential improvements on
computation speed using quantum mechanics, is the one which greatly
hinders immediante technical implementation. Entanglement between
different subsystems is essential for the production of the states
used in information processing; at the same time it prevents these
qubits to be completely isolated from its environment, producing
undesired entanglement with the environmental degrees of freedom. The deleterious
effect of this coupling is usually called decoherence. Therefore much
effort has been devoted to finding ways around decoherence in quantum
computation, such as error correcting
codes\cite{codes}, dynamical decoupling\cite{dynadeco} and computation
in decoherence free 
subspaces\cite{DFS}. Experimental observations of  decoherence free evolution
have been reported.\cite{Science1,Science2} Many physical
implementations have been proposed including cavity QED\cite{CQED}, ions
traps \cite{Ion}, nuclear magnetic resonance \cite{NMR} and
semiconductor quantum dots\cite{QDots}. From the theoretical point of
view, recent work has been mainly on proving the existence of DF subspaces. Much 
less work, however has been devoted to the dynamics. In general those
models are highly idealized, and it is of interest at least to relax some of the
stringent conditions, e.g. degeneracy of the free modes, and calculate
such effects on the time evolution of the system. This could be
considered as an important step towards realistic implementations of
the models.

In the present work we consider the case of N independent oscillators
linearly coupled to a single environment, and show that the existence
of  strict decoherence free subspaces can only be obtained under the
following two conditions: degeneracy of the oscillators, and
separability of the coupling with the environment. Both the exact form of
the spectral density and the temperature of the environment are
immaterial in what concerns the existence of DFS. A formal equation
for the dynamics of the reduced density matrix of the oscillators is
derived which resembles the corresponding master equation, but with
time dependent coefficients. For two independent oscillators we solve
the dynamics of the reduced density and construct the decoherence free
subspace. Moreover under Markov approximation, and zero temperature,
we are able to give an analytical expression for the time evolution of
the system of interest, in the general case (i.e. non degenerate modes
and nonseparable coupling). We study the effect of a slight breaking
of the degeneracy and separability and show that the dynamics will in
general produce a long lived and a short live components. The time
scale for the duration of these components is derived in terms of the
appropriate parameters. 

\section{The model}

Let us consider a collection of N identical harmonic oscillators 
linearly coupled
to a single environment, modelled as a large set of harmonic
oscillators,
\[
{\bf \hat{H}}=\omega \sum_{i=1}^N  {\bf \hat{a}_i^\dagger}{\bf \hat{a}_i}
+
\sum_k \omega_k{\bf \hat{b}_k^\dagger}{\bf \hat{b}_k}
+ \sum_{i,k} (g_{ik}^* {\bf \hat{a}_i^\dagger}{\bf \hat{b}_k} 
+ g_{ik} {\bf \hat{a}_i}{\bf \hat{b}_k^\dagger})
\]
If the coupling of the N oscillators to the environment does not
depend on the particular oscillator, apart from an overall factor, 
the coupling constants $g_{ik}$ can be written as a product $G_i
D_k$. In this case it is immediate to verify that there is a single
collective mode which will remain coupled to the environment, while
the other N-1 collective modes form a decoherent free (infinite)
subspace.\cite{DFS} The collective mode can be easily visualized
if we rewrite the hamiltonian as follows
\[
{\bf \hat{H}}=
\omega \sum_{i=1}^N  {\bf {\hat a}_i^\dagger}{\bf \hat{a}_i}
+
\sum_k \omega_k {\bf \hat{b}_k^\dagger \hat{b}_k}
+ \sum_{k} \left(D_{k}^* (\sum_i G_i^* {\bf \hat{a}_i^\dagger}){\bf \hat{b}_k} 
+ D_{k}(\sum_i G_i {\bf \hat{a}_i}) {\bf \hat{b}_k^\dagger}\right),
\]
where it is clear that the collective mode coupled to the environment
has the following creation and annihilation operators
\[
{\bf \hat{ A}^\dagger_1} = \frac{\sum_i G_i^*{\bf \hat{a}_i^\dagger}}{\sum_i |G_i|^2}, \quad
{\bf \hat{A}_1} = \frac{\sum_i G_i{\bf \hat{ a}_i}}{\sum_i |G_i|^2}.
\]
This can be viewed as a N-dimensional rotation in the space of the 
original creation and annihilation operators, which maps the set ${\bf \hat{a}_i}$
onto the set ${\bf \hat{A}_i}$ in terms of which the hamiltonian is rewritten as
\[
{\bf \hat{H}}=\omega \sum_{i=1}^N  {\bf \hat{A}_i^\dagger}{\bf \hat{ A}_i}
+
\sum_k \omega_k{\bf \hat{ b}_k^\dagger \hat{ b}_k}
+  {\bf \hat{A}_1^\dagger}\sum_{k} c_{k}{\bf \hat{ b}_k} 
+ {\bf \hat{ A}_1}\sum_{k} c_{k}^*{\bf \hat{ b}_k^\dagger},
\]
with $c_k = \sum_i |G_i|^2 D_k^*$. The rotation leading to the equation above
defines the relation between the new and old operators. This rotation
is not unique except in the case of two identical operators.
We show next that in contrast with other commonly studied models, in
the present one  we are able to introduce both decoherence and
dissipation, for arbitrary environmental spectral densities 
(ohmic, subohmic, etc) and temperatures. A long but straighforward
procedure\cite{alamos} leads to the following exact generalized master
equation
\begin{equation} \label{Eq:Master}
\begin{split}
\frac{d{\bf \hat{\rho}}}{dt} = &
\frac{1}{i\hbar}\left[{\bf \hat{H}_0},{\bf \hat{\rho}} \right]
+(\lambda+\epsilon)(2 {\bf \hat{A}_1\cdot \hat{A}_1^\dagger}-{\bf \hat{A}_1^\dagger \hat{A}_1\cdot} 
-{\bf \cdot \hat{ A}_1^\dagger\hat{ A}_1}){\bf \hat{\rho}}\\
& +\epsilon(2{\bf \hat{ A}_1^\dagger\cdot \hat{A}_1}-{\bf \hat{A}_1 
\hat {A}_1^\dagger\cdot} 
-{\bf \cdot\hat{ A}_1 \hat{A}_1^\dagger}){\bf \hat{\rho}},
\end{split}
\end{equation}
where 
\[{\bf \hat{ H}_0}=\hbar\omega \sum_{i=2}^N{\bf \hat{ A}_i^\dagger\hat{ A}_i}+
\hbar(\omega+\delta){\bf \hat{A}_1^\dagger\hat{ A}_1}. 
\]
The real functions $\lambda,\delta,\epsilon$ are implicitly defined in
terms of the  auxiliary function $\eta(t)$
\begin{equation}\label{eta}
\eta(t)=\exp\left(
-\int_0^t \lambda(t') dt'-iwt -i\int_0^t \delta(t')dt'
\right)
\end{equation}
which satisfies the 
integrodifferential equation
\begin{equation} \label{Eq:Integrodif}
\dot{\eta}+i\omega \eta + \int_0^t d\tau
\sum_k |c_k|^2 {\rm e}^{i\omega_k(t-\tau)} \eta(\tau)=0,
\end{equation}
subject to the initial condition $\eta(0)=1$.
Moreover we have
\[
\epsilon(t) =  \frac{|\eta(t)|^2}{2 }\frac{d}{dt}
\left( \sum_k
\frac {|c_k|^2 n_k (\beta)}{|\eta(t)|^2} 
\left|
\int_0^t d\tau e^{-i\omega_k (t-\tau)} \eta(\tau)
\right|^2 \right).
\]
where $n_k(\beta)$
is the mean excitation number for the k-th mode of the environment
at inverse temperature $\beta= 1/k_B T$. If the usual Born-Markov
approximations hold, then $\delta(t) =0$, $\lambda(t)= \sum_i k_i$,
and $\epsilon=\sum_i k_i\bar{n}$, where $k_i$  characterize the
markovian evolution when the only the i-th original oscillator is 
coupled to the bath, and $\bar{n}$ is the environment mean number
of thermal excitations. In this case the master equation 
simplifies to
\begin{equation} \label{Eq:BMMaster}
\begin{split}
\frac{d{\bf \hat{\rho}}}{dt} = &
-i\omega\sum_{i=1}^N \left[{\bf \hat{ A}_i^\dagger\hat{ A}_i},{\bf \hat{\rho}} \right]
+\sum_{i=1}^N k_i (\bar{n}+1)(2{\bf \hat{ A}_1\cdot \hat{A}_1^\dagger}-{\bf \hat{A}_1^\dagger \hat{A}_1\cdot} 
-{\bf \cdot \hat{A}_1^\dagger \hat{A}_1}){\bf \hat{\rho}}\\
& +\sum_{i=1}^N k_i \bar{n}(2{\bf \hat{ A}_1^\dagger\cdot\hat{ A}_1}-{\bf \hat{A}_1 
\hat{A}_1^\dagger\cdot} 
-{\bf \cdot \hat{A}_1\hat{ A}_1^\dagger}){\bf \hat{\rho}}.
\end{split}
\end{equation}
Note that the collective mode which couples to the environmental
degrees of freedom 
decoheres much faster than any of the individual oscillators.

\subsection{Two Identical Oscillators}

For the sake of definiteness we 
restrict ourselves to the case of two identical oscillators.
Let us see that in this case the collective mode ${\bf \hat{A}_1}$ (${\bf \hat{A}_2}$) which
couples (decouples) to the environment is given by 
\[
{\bf \hat{A}_1^{(\dagger)}} = \cos (\theta) {\bf \hat{a}_1^{(\dagger)}} 
                  + \sin(\theta) {\bf \hat{a}_2^{(\dagger)}},\quad
{\bf \hat{A}_2^{(\dagger)}} = -\sin (\theta) {\bf \hat{a}_1^{(\dagger)}} 
                  + \cos(\theta) {\bf \hat{a}_2^{(\dagger)}},
\] 
with $\tan(\theta)=G_2/G_1$, which reduces to  $\tan(\theta)=\sqrt{k_2/k_1}$
when the markovian approximation is valid. Let us see that the
transformation above can be implemented as using an appsropriate
rotation operator.
The generalized master
equation (\ref{Eq:Master}) for two identical oscillator can be exactly
solved using the evolution superoperator $\mathcal{U}(t)$
\begin{equation} \label{SuperU}
\mathcal{U}(t) = 
e^{-i\theta[i({\bf \hat{a}_1 \hat{a}_2^\dagger}-{\bf \hat{a}_2
\hat{a}_1^\dagger}),{\bf \cdot}]} {\rm
e}^{-iwt[{\bf \hat{a}_2^\dagger \hat{a}_2} ,{\bf \cdot}]}
 v {\rm e}^{(1-v) {\bf \hat{a}^\dagger_1 \cdot \hat{a}_1}} {\rm e}^{x {\bf \hat{a}^\dagger_1 \hat{a}_1 \cdot}} 
{\rm e}^{x^*{\bf \cdot \hat{a}^\dagger_1 \hat{a}_1}} 
{\rm e}^{z {\bf \hat{a}_1\cdot \hat{a}^\dagger_1}}
e^{i\theta[i({\bf \hat{a}_1 \hat{a}_2^\dagger}-{\bf \hat{a}_2
\hat{a}_1^\dagger}),{\bf \cdot}]}
\end{equation}
where the coefficients $v(t)$, $x(t)$ and $z(t)$ can be
given in terms of the functions $\eta(t)$, eq. (\ref{eta}), and
$\mathcal{N}(t)$,
\[
 {\cal N}(t)=
\int_0^t d\tau \epsilon(\tau)
\left|\frac{\eta(\tau)}{\eta(t)}\right|^2,
\]
as follows 
\begin{equation} \nonumber
v(t)  =  \frac{1}{1+{\cal N}(t)}, \quad
x(t)  =  \ln \frac{\eta(t)}{\sqrt{1+{\cal N}(t)}}, 
\quad z(t)  =  1-\frac{\left|\eta(t)\right|^{-2}}{1+{\cal N}(t)}.
\end{equation}
In the markovian limit the preceding formulas reduce to
\[
v=\frac{1}{1+\bar{n} (1-e^{-2(k_1+k_2)t})},\quad 
x= \ln \frac{e^{(-i\omega-k_1-k_2) t}}{\sqrt{1+\bar{n}(1-e^{-2( k_1 +k_2)t})}},\quad
z=\frac{(\bar{n}+1)(1-e^{-2(k_1+k_2)t})}{1+\bar{n}(1-e^{-2(k_1+k_2)t)})}.
\]
In eq. (\ref{SuperU}) the first and last terms of the rhs correspond
to the rotation which leads to the coupled and uncoupled collective
modes.

Since the second collective mode is effectively decoupled from the 
environment, any density operator in the Hilbert space of this mode, 
times the asymptotic density operator of the coupled collective mode,
provided it exists, will experience a unitary evolution. For simplicity
we will further on restrict ourselves to the zero temperature case.
Any density operator of the form

\[
\begin{split}
{\bf \hat{\rho}} & = \sum_{n,m} \frac{{\bf \hat{\rho}}_{n,m}}{n!m!} 
(-{\bf \hat{a}_1^\dagger} \sin(\theta)+{\bf \hat{a}_2^\dagger} \cos(\theta))^n
\ket{0,0}\bra{0,0}
(-{\bf \hat{a}_1} \sin(\theta)+{\bf \hat{a}_2} \cos(\theta))^m
\\ & =
\sum_{n,m,n_1,m_1} {\bf \hat{\rho}}_{n,m}
\frac{\sqrt{n!m!}(-\sin\theta)^{n+m-n_1-m_1}(\cos\theta)^{n_1+m_1}}
{\sqrt{(n-n_1)!n_1!(m-m_1)!m_1!}}
\\ & \qquad \qquad \quad \times 
\ket{n_1,n-n_1}\bra{m_1,m-m_1},
\end{split}
\]
will be protected against dissipation and decoherence. In fact,
applying the evolution superoperator to an initial density matrix of
this form, we obtain

\[
\begin{split}
{\bf \hat{\rho}}(t) &  =
\sum_{n,m,n_1,m_1} e^{-i\omega t (n-m)}{\bf \hat{\rho}}_{n,m}
\frac{\sqrt{n!m!}(-\sin\theta)^{n+m-n_1-m_1}(\cos\theta)^{n_1+m_1}}
{\sqrt{(n-n_1)!n_1!(m-m_1)!m_1!}}
\\ & \qquad \qquad\qquad\qquad \qquad \times 
\ket{n_1,n-n_1}\bra{m_1,m-m_1},
\end{split}
\]

Now, we use the evolution superoperator on the initial operator density 
\begin{equation}\label{initialcondition}
{\bf \hat{\rho}}(0)=(\cos(\alpha)\ket{1,0}+\sin(\alpha)e^{i\phi}\ket{0,1})
(\cos(\alpha)\bra{1,0}+\sin(\alpha)e^{-i\phi}\bra{0,1}),
\end{equation}
to obtain the time dependent evolution, which can be given in closed
form and corresponds to the limit where the frequencies are degenerate
and coupling is separable, of the formulas given in the next section.
For now we show the asymptotic limit,
\[
{\bf \hat{\rho}}(t\rightarrow\infty)  = P \ket{\psi}\bra{\psi}
+(1-P)\ket{0,0}\bra{0,0}
\]
where the probability to go to the state $\ket{\psi}$,
\[
\ket{\psi} =\frac{\sqrt{k_2} \ket{1,0} -\sqrt{k_1} \ket{0,1}}{\sqrt{k_1+k_2}}
\]
is given by
\[
P =
\left|\left| 
\frac{\sqrt{k_2}\cos(\alpha)-\sqrt{k_1}e^{i\phi}\sin(\alpha)}
     {\sqrt{k_1+k_2}}
\right|\right|^2,
\]
and the probability to go to the joint ground state is $1-P$.

Observe that varying $\alpha$ and $\phi$ we can go from total preservation
to total leakage. For example, if we set $\tan(\alpha) =
\sqrt{k_1/k_2}$, and $\phi=0$ then the full state will leak to the
ground state $\ket{0,0}$. On the other hand, if we set $\tan(\alpha) =
-\sqrt{k_2/k_1}$, and $\phi=0$ then the initial state will be
exactly equal to $\ket{\psi}$, and will persist at all
times with probability 1. All other combinations will go to the
asymptotic state $\ket{\psi}$ with probability $P$ and to the
ground state $\ket{0,0}$ with probability $1-P$. The asymptotic
fidelity, $F(\infty)$, which is the overlap between the initial and final density
matrices, is given by
\[
F(\infty) = \left|\left| 
\frac{(\sqrt{k_2}\cos(\alpha)-\sqrt{k_1}e^{i\phi}\sin(\alpha))
(\sqrt{k_2}\cos(\alpha)-\sqrt{k_1}e^{-i\phi}\sin(\alpha))}
{k_1+k_2}\right|\right|^2
\]

\section{Effects of more Realistic Modelling}

We remark that the results above were obtained under a number of
assumptions, which will be relaxed below. Notice that the use
of the rotating wave approximation (RWA) is not essential in obtaining
the decoupled mode: any interaction linear in the field operators
would be as good (provided the other assumptions hold). Had we chosen
an interaction linear in the identical oscillators but nonlinear on
the environmental operators we would have obtained also a decoupled
collective mode. In these cases, however, the complication would be
only of technical nature leading to (much) more complex dynamics. 

Another important hypothesis to obtain DFS is that of identical
frequencies of the original main oscillators. Of course, any
interaction between them would destroy the symmetry upon which the
existence of DFS rests. On the other hand, we have assumed that the
oscillator-environment coupling satisfies $g_{ik} = G_i D_k$, which
amounts to a separable coupling. It is not an easy task to find
realizations of such interactions in nature given its nonlocal
character. However, it might be a good approximation in special
circumstances, as e.g. optical cavities. A particular consequence of
the separability hypothesis can be seen writing the master equation,
in the zero temperature limit, 
in terms of the original oscillators (with different frequencies
for generality)
\begin{eqnarray}\label{MasterdaSonia}
{\cal L}_0 & = &\ad\aponto\left(-i\omega_{1} 
- {k_1}\right)+\pad{\bf \hat{a}}\left( i\omega_{1}-{k_1}\right)
+2{k_1}\aponto\ad\nonumber\\
& & +\bd\bp\left(-i\omega_2-{k_{2}}\right)
+\pbd{\bf \hat{a}_{2}}\left( i\omega_{2}-{k_{2}}\right)
+2{k_{2}}\bp\bd\nonumber\\
&&+{k_{3}}\left(2\aponto\bd-\ad\bp-\pad{\bf \hat{a}_{2}}\right)
+{k_{3}}\left(2\bp\ad-{\bf \hat{a}}\bdp-\aponto\bd\right).
\end{eqnarray}
Notice that we have introduced a new quantity $k_3$, which is the
environment induced coupling. When the separability condition holds the
three dissipation constants, $k_1,k_2,k_3$ satisfy the relationship
$k_3^2 = k_1 k_2$. Since these constants are given by the expressions
\begin{eqnarray}\label{hrb}
\left|g_{1k}\right|^{2}
\approx\frac{k_{1}}{\pi},
\quad \left|g_{2k}\right|^{2}\approx\frac{k_{2}}{\pi},
\quad
g_{1k}g_{2k}^{*}(\omega)\approx\frac{k_{3}}{\pi}
\end{eqnarray} 
Schwarz inequality implies immediately that $|k_3|^2 \leq k_1 k_2$ in the
general (nonseparable) case. In what follows we present the general
solution to eq. (\ref{MasterdaSonia}) and study, in particular, the dynamics
for parameters which do not quite satisfy the condition for DFS, but
both the frequency degeneracy and coupling separability are almost
fulfilled. 

The evolution superoperator for eq.(\ref{MasterdaSonia}) is given by\cite{Sonia}
\begin{eqnarray}\label{rot}
{\bf \hat{\rho}}(t)
& = & e^{\cjat\aponto\ad}e^{\cjbt\bp\bd}e^{\czt\bp\ad}
e^{\cczt\aponto\bd}
e^{\cqt{\bf \hat{a}_{1}}\bdp}
e^{\ccqt\pad{\bf \hat{a}_{2}}}e^{\cmbt\bd\bp}
e^{\ccmbt\pbd{\bf \hat{a}_{2}}}
\odot\nonumber\\
&&\odot  e^{\cmat\ad\aponto}
e^{\ccmat\pad{\bf \hat{a}_{1}}}
e^{\cqt\ad\bp}
e^{\ccqt\aponto]
e\bd}{\bf \hat{\rho}}_0
\end{eqnarray}
where
\begin{eqnarray*}
&&R=\frac{ k_2+
k_1}{2}+\frac{i\left(\omega_2+\omega_1\right)}{2},\quad c= k_2- k_1
+i\left(\omega_2-\omega_1\right),\quad 
r=\sqrt{c^2+4k_3^{2}},\quad \Delta_{\pm}=
c\pm r \\
&&\cqt= 2k_3\left(1- e^{r\;t}\right)\left(
\Delta_{+}e^{r\;t}-\Delta_{-}\right)^{-1}\quad
{\textnormal{for}}\quad r\not=0 \\
&&e^{\cmat}=\frac{e^{-R\;t}}{2r}
e^{-\frac{r\;t}{2}}\left(\Delta_{+}e^{r\;t}-\Delta_{-}\right),
\quad e^{\cmbt}=e^{-2R\;t}e^{-\cmat} \\
&&\cjbt=\left(1+|\cqt|^{2}\right)
\left(\left|e^{\cmbt}\right|^{-2}\right)-1\\
&&\cjat=\left|e^{-\cmat}+\cqt^2 e^{-\cmbt}\right|^{2}
+\left(|\cqt|^2\right)\left(\left|
e^{\cmbt}\right|^{-2}\right)-1\\
&&\czt=-\cqt e^{-\left(\ccmat+\cmbt\right)}-\ccqt
\left( 1+|\cqt|^{2}\right)\left|e^{\cmbt}\right|^{-2}.
\end{eqnarray*}
Note that the neither the separability nor the degeneracy conditions
have been used so far in obtaining the solution.
For the sake of comparison we use the same initial condition of the
precedent section (eq. (\ref{initialcondition})). In the general case
its time evolution is given by
\[
{\bf \hat{\rho}}(t)  = P(t) \ket{\psi(t)}\bra{\psi(t)}
+(1-P(t))\ket{0,0}\bra{0,0}
\]
where the probability to go to the state $\ket{\psi(t)}$,
\[
\ket{\psi(t)} =
\frac{(\cos(\theta)M_-(t)+\sin(\theta)e^{i\phi} Q(t)) \ket{1,0} 
+(\sin(\theta)e^{i\phi} M_+(t)+\cos(\theta)Q(t))\ket{0,1}}{\sqrt{P(t)}}
\]
is given by
\[
P(t) =
\left|\left| 
\cos(\theta)M_-(t)+\sin(\theta)e^{i\phi} Q(t)
\right|\right|^2+
\left|\left| 
\sin(\theta)e^{i\phi} M_+(t)+\cos(\theta)Q(t)
\right|\right|^2,
\]
and the probability to go to the joint ground state is $1-P(t)$. The
functions $M_{\pm}(t)$ and $Q(t)$ are given by
\[
M_{\pm}(t) = \frac{e^{-R t}}{2} 
\left( 
e^{- r t/2} (1\mp\frac{c}{r}) +e^{r t/2} (1\pm\frac{c}{r})
\right) , \quad
Q(t) = \frac{k_3}{r} e^{-R t}\left(e^{- r t/2}  - e^{r t/2}\right).
\]
Now, we assume slight deviations from degeneracy and separability,
that is, $\omega_1 = \omega-\delta \omega$, $\omega_2 = \omega+\delta \omega$, $k_3 =
\sqrt{k_1 k_2} -\delta k$, with $\delta \omega\ll \omega$, $\delta k\ll \sqrt{k_1 k_2}$.
Then, the state $\ket{\psi(t)}$ can be
approximated as 
\[
\ket{\psi(t)} = e^{-i\omega t}
\times \left(
\frac{\zeta_1\ket{1,0} +\xi_1\ket{0,1}}{\sqrt{P(t)}} e^{-(k_1+k_2)t}
      +\frac{\zeta_2 \ket{1,0} 
      +\xi_2 \ket{0,1}}{\sqrt{P(t)}} 
e^{-\frac{2\delta k \sqrt{k_1 k_2}}{k_1+k_2}t}
\right),
\]
where $\zeta_i,\xi_i$ do not have any temporal dependence and are
given by
\begin{eqnarray}
 \zeta_{1\choose 2} & = &
\frac{(k_{1 \choose 2}\mp i\delta \omega)\cos(\alpha)
\pm\sqrt{k_1k_2}\sin(\alpha)e^{i\phi}}{k_1+k_2},
\\
\xi_{1\choose 2} & = &
\frac{(k_{2\choose 1}\pm i\delta \omega) e^{i\phi}\sin(\alpha)
	\pm\sqrt{k_1k_2}\cos(\alpha)}{k_1+k_2}.
\end{eqnarray}
In the general case it is not possible to find initial conditions
which are completely decoherence free. Nor it is possible to find to
orthogonal subspaces with very different characters in what
decoherence is concerned. However, we can choose the initial condition
as to have a minimal component either in a strong decoherence (SD) or in a
weak decoherence (WD) subspaces, by choosing, e.g.
\[
tan(\alpha)_{{\mbox{\tiny{SD}}}\choose{\mbox{\tiny{WD}}}} =
\pm \sqrt{\frac{k_{2\choose 1}}{k_{1\choose 2}}},
\quad \phi_{{\mbox{\tiny{SD}}}\choose{\mbox{\tiny{WD}}}} = \pm \frac{\delta\omega}{k},
\]
where $k$ is some average dissipation constant. The corresponding
subspaces, apart from a phase, can be written as 
\begin{equation}\label{psiwd}
\ket{\psi_{{\mbox{\tiny{SD}}}\choose{\mbox{\tiny{WD}}}}} =
\frac{1}{\sqrt{k_1+k_2}}
\left(\sqrt{k_{1\choose 2}}\ket{1,0} 
\pm\sqrt{k_{2\choose 1}}e^{\pm i \delta \omega/k}\ket{0,1} \right).
\end{equation}
If $\delta \omega\ll k_1, k_2$ the phase can be ignored. Moreover, if
$k_1=k_2$ then we have $k=k_1=k_2$ and the phase can be unambigously
determined. 
The weak decoherence wavefunction defines a space which is robust
against decoherence with typical decoherence time scale of the order
\[
\tau_{WD}=\frac{k_1+k_2}{2\delta k\sqrt{k_1 k_2}},
\]
 which is to be compared
with the strong decoherence time associated to $\ket{\psi_{SD}}$ given
by 
\[
\tau_{SD}=\frac{1}{k_1+k_2}.
\]
If $k_1$ and $k_2$ are of approximately the same order ($k$), then the weak
decoherence time is about $k/\delta k$ times greater than the strong
decoherence time. In all other cases (different from $\delta \omega=0$ or
$k_1=k_2$ one always has, for any initial condition (of the form
(\ref{initialcondition})), a term which decays slowly and one which decays rapidly
comparatively. However, the forms given above may help one to choose
the initial condition in such a way that the time evolution is robust
against decoherence (it should be close to form (\ref{psiwd}) with
$k=(k_1+k_2)/2$).

Had we chosen other Fock subspaces (for example those of $m$ total
excitations), we would have obtained also an analytical solution
showing both slowly and fast decaying components. Of course, the
expressions get more involved.

\section{Concluding Remarks}

We have studied decoherence and dissipation free subspaces for systems
involving N identical harmonic oscillators linearly coupled to a
single environment. The conditions for the existence of such decoherent
free subspaces are too stringent, and hardly implementable in
practice. Therefore, we studied the dynamics of such systems when
these conditions are slightly broken. This may serve as a guide for
preparation of initial conditions wich are fairly robust against
decoherence. To the extent that this model is adequate for the
description of modes in a QED cavity, it would be possible to use it
to generate good quantum memories in the presence of dissipation and
decoherence.

\section{Acknowledgements}
We gratefully acknowledge comments from A. N. Salgueiro.
This work was partly funded by FAPESP, CNPq and PRONEX (Brazil), 
and Colciencias, DINAIN (Colombia). K.M.F.R. gratefully acknowledges 
the Instituto de F\'\i sica, Universidade de S\~ao Paulo, for their
hospitality and PRONEX for partial support.

\end{document}